\documentclass[fleqn,twoside]{article}
\usepackage{espcrc2}
\usepackage{graphicx}
\def\beq{\begin{equation}}
\def\eeq{\end{equation}}
\def\bqry{\begin{eqnarray}}
\def\eqry{\end{eqnarray}}

\def\qbar{{\overline{q}}}
\def\sbar{{\overline{s}}}

\def\tq{{\tilde{q}}}
\def\ts{{\tilde{s}}}
\def\td{{\tilde{d}}}
\def\tu{{\tilde{u}}}
\def\tqbar{{\overline{\tq}}}

\def\str{{\rm str\;}}
\def\Nhat{{\hat N}}

\newcommand{\AmS}{{\protect\the\textfont2
  A\kern-.1667em\lower.5ex\hbox{M}\kern-.125emS}}

\title{Unfactorizing Polychromatic Penguins.}

\author{Santi Peris\address{Grup de Fisica Teorica and IFAE,
Universitat Autonoma de Barcelona, 08193 Barcelona, Spain}}

\begin{document}

\begin{abstract}

Firstly, I report on a recent work --carried out with T. Hambye and E. de Rafael-- on how
to use the large-$N_c$ expansion to calculate unfactorized contributions from the strong
penguin operators, and their impact on observables such as $\epsilon'/\epsilon$ and the
$\Delta I=1/2$ rule. Secondly, based on work done with M. Golterman, I explain how this
calculation allows one to predict some rather dramatic consequences for quenched QCD.
This may help explain the present discrepancy between lattice and experimental results
for $\epsilon'/\epsilon$. The emphasis of this article is put on the explanation of the
method of calculation used, which is fully analytic. This allows one to build some
intuition and understand the role played by the different hadronic scales in determining
the size of the different contributions.

\vspace{1pc}
\end{abstract}

\maketitle

Due to the disparity of scales between the kaon and the W masses, the decay process
$K\rightarrow \pi\pi$ has very large logarithms, such as  $\log M_W/M_K$. The technique
of Effective Field Theory allows one to resum all these large logarithms as the
renormalization group running of certain new couplings ( the so-called Wilson
coefficients). These coefficients appear in the Effective Lagrangian where the $W$ and
all the other heavy particles of the Standard Model are no longer present, i.e. they have
been  ``integrated out''. Therefore, in this Effective Lagrangian only the u,d, s quarks
and the gluon fields appear explicitly, the effect of the heavy particles being encoded
in a set of 10 Wilson coefficients and equal number of four-quark operators\footnote{See,
e.g., ref. \cite{Buras} for the full list of $\Delta S=1$ operators.} , i.e.
\begin{equation}\label{one}
    \mathcal{L}_{eff}= \sum_{AB} C_{AB}(\mu)\ \mathcal{Q}_{AB}(\mu)\ ,
\end{equation}
where $\mathcal{Q}_{AB}(\mu)=\overline{q}\Gamma_A q(x)\ \overline{q}\Gamma_B q(x)$ and
 $\Gamma_{A,B}$ are matrices in Dirac space (color indices are suppressed).
This Effective Lagrangian, $\mathcal{L}_{eff}$ changes the short-distance properties of
Green's functions with respect to the full Standard Model, where the heavy particles were
explicit fields in the Lagrangian. A set of matching conditions ensure that the physics
does not change, however. These matching conditions define Wilson coefficients and
operators and make the whole construction meaningful so that the scheme dependence
appearing in the Wilson coefficients cancels that appearing in the four-quark operators'
matrix elements.

In perturbation theory all this is very well understood. Two groups have been able to
compute the Effective Lagrangian even to two loops\cite{anomdim}. However, perturbation
theory may be a sensible approximation down to the integration of the charm quark,  but
certainly not to describe the matrix elements of the kaon. In fact, since the kaon is a
(pseudo) Goldstone boson, its dynamics is appropriately described in terms of yet another
Effective Field Theory. In the $\Delta S=1$ sector, this is given by the Chiral
Lagrangian\footnote{In the large-$N_c$ expansion one has to include the $\eta'$ in a
nonet.}:
\begin{eqnarray}\label{ewcl}
\mathcal{L}_{eff}^{\Delta S=1}&&\!\!\!\!\!\!\!\!\!=-\frac{G_F}{\sqrt{2}}\ V_{ud}V_{us}^*\\
&&\!\!\!\!\!\!\!\!\!\!\!\!\!\!\!\!\!\!\!\!\!\!\!\!\!\!
\left[g_{\underline{8}}\mathcal{L}_{8}+ g_{\underline{27}}\mathcal{L}_{27}+e^2
g_{ew}\mathrm{Tr}\left(U\lambda_ {L}^{(32)}U^{\dag}Q_R \right) \right]\nonumber \,,
\end{eqnarray}
where
\begin{eqnarray}
\mathcal{L}_8&=&\sum_{i=1,2,3}(\mathcal{L}_{\mu})_{2i}\ (\mathcal{L}^{\mu})_{i3}\quad
\mathrm{and}\\
 \quad \mathcal{L}_{27}&=&\frac{2}{3}(\mathcal{L}_{\mu})_{21}\
(\mathcal{L}^{\mu})_{13}+ (\mathcal{L}_{\mu})_{23}\ (\mathcal{L}^{\mu})_{11}\, ,
\end{eqnarray}
 with
 \begin{equation}
 \mathcal{L}_{\mu}=-iF_{0}^2\ U(x)^{\dagger}D_{\mu} U(x)\ ,\
 \lambda_{L}^{(32)}=\delta_{i3}\delta_{j2}\end{equation}
 and
 \begin{equation}
Q_{L}=Q_{R}=Q=\mbox{\rm diag.}(2/3,-1/3,-1/3)\,. \end{equation} The pion decay coupling
constant $F_0$ is in the chiral limit, i.e. $F_0\simeq 87$~MeV. The matrix field $U$
contains the Goldstone fields of the corresponding spontaneously broken chiral symmetry
of QCD, and $D_{\mu}U$ stands for the covariant derivative
$D_{\mu}U\!=\!\partial_{\mu}U-ir_{\mu}U+iUl_{\mu}$  in the presence of external chiral
sources $l_{\mu}$ and $r_{\mu}$ of left-- and right--handed currents. The Lagrangian
(\ref{ewcl}) can be thought of as the one obtained from (\ref{one}) after integrating out
all the hadrons but the Goldstone octet. How does one go about matching the two Effective
Field Theories (\ref{one}) and (\ref{ewcl})? \footnote{Our approach to this problem has
also been explained in \cite{Tempe}.} At this point we find it very convenient to use the
large-$N_c$ expansion to organize the calculation because, being a systematic expansion
in QCD,  it can be carried out either with quarks and gluons or with mesons as degrees of
freedom. So, in a way, the large-$N_c$ expansion plays the role of a dictionary between
these two languages.

 In the following I will concentrate on how to impose the matching condition for the
case of the strong penguin operators
\begin{eqnarray}\label{penguins}
  \mathcal{Q}_6 &=& -8\sum_{q=u,d,s}
  \left(\overline{s}_Lq_R\right)\left(\overline{q}_Rd_L\right)\ ,\ \mathrm{and}\nonumber \\
  \mathcal{Q}_4 &=& 4 \sum_{q=u,d,s}
  \left(\overline{s}_L\gamma^{\mu}q_L\right)\left(\overline{q}_L\gamma_{\mu}d_L\right)\ ,\
\end{eqnarray}
where $q_{L,R}=\frac{1}{2}(1\mp\gamma_5)q$ and sum over color indices within brackets is
understood. Notice that, since four-quark operators mix among themselves, it makes no
sense to consider the contribution from one of them in isolation. However, as it turns
out, if we are willing to restrict ourselves to the leading  logarithmic approximation
and only to those subleading contributions in $1/N_c$ which are enhanced by the number of
flavors $n_F$, there is an important simplification since the $\mathcal{Q}_{4,6}$ system
closes and no other operator needs be considered through mixing. I will adopt this
simplification in the following.

Firstly, let us define the constant $g_{\underline{8}}$ in the Lagrangian \ref{ewcl} as
the coupling governing the non-diagonal ``mass term'' $r_{\overline{d}q}^{\alpha}
r_{\overline{q}s}^{\beta}\ g_{\alpha\beta}$ for any flavor $q=u,d,s$.  For definiteness,
we shall take $q=u$. The matching condition stating that the same term be obtained with
the Lagrangian (\ref{one}) reads\cite{HPdR}

\begin{eqnarray}\label{matching}
&&\!\!\!\!\!\!\!\!\!\!\!\!\!\!\!\!\!g_{\underline{8}}\vert_{Q_4,Q_6} =
C_{6}(\mu)\Bigg\{\frac{\!-16L_{5}\langle
\overline{\psi}\psi\rangle^2}{F_{0}^6} \nonumber \\
&&\!\!\!\!\!\! \!\!\!\!\! + \frac{8n_f}{16\pi^2 F_{0}^4}\ \int_{0}^{\infty}\!\!\!dQ^2
Q^{2}\ \mathcal{W}_{DGRR}(Q^2)
\Bigg\}_{\overline{\mathrm{MS}}} \\
 & \!\!\!\!\!\!\!\!\!\!\!+ &\!\!\!\! \!\!\!\!
 C_{4}(\mu)\Bigg\{1\!-\!\frac{4n_f}{16\pi^2 F_{0}^4}\int_{0}^{\infty}\!\!\!\!\! dQ^2 Q^{2}\ \mathcal{W}_{LLRR}(Q^2)
\Bigg\}_{\overline{\mathrm{MS}}}\ .\nonumber
\end{eqnarray}
The subscript $\overline{\mathrm{MS}}$ reminds one that these integrals are UV divergent
and have to be regularized and renormalized, using the same scheme as for the Wilson
coefficients $C_{4,6}$. The parameters $\langle \overline{\psi}\psi\rangle$ and $L_5$ are
also  renormalized accordingly. In Eq. (\ref{matching}) all the unfactorized
contributions, of $\mathcal{O}(n_F/N_c)$, are contained in the terms proportional to the
functions $\mathcal{W}_{DGRR}$ and $\mathcal{W}_{LLRR}$. The terms proportional to $L_5$
and unity correspond to the factorized contribution from $Q_6$ and $Q_4$ --respectively--
and, formally, are of $\mathcal{O}(N_c^0)$.

The functions $\mathcal{W}_{DGRR}$ and $\mathcal{W}_{LLRR}$ are defined through the
connected four-point Green's functions
\begin{eqnarray}\label{green}
  &&\!\!\!\!\!\!\!\!\!\mathcal{W}_{DGRR}^{~~~~\alpha\beta}(q)\!=
  i^3  \int  d^4x\ d^4y\ d^4z\
e^{i q.x}\nonumber \\
&& \!\!\!\langle 0\vert T\{ D_{\bar{s}q}(x)G_{\bar{q}d}(0)
R_{\bar{d}u}^{\alpha}(y)R_{\bar{u}s}^{\beta}(z)\}\vert 0\rangle_{\mbox{\rm\tiny
conn.}}\ ,\nonumber \\
&&\!\!\!\!\!\!\!\!\!\!\mathcal{W}_{LLRR}^{\,\mu\,\mu\,\alpha\,\beta}(q)
= i^3 \int d^4x\ d^4y\ d^4z\ e^{iq\cdot x} \nonumber \\
&&\!\!\!\langle 0\vert T\{ L_{\bar{s}q}^{\mu}(x)L^{\bar{q}d}_{\mu}(0)
R_{\bar{d}u}^{\alpha}(y)R_{\bar{u}s}^{\beta}(z)\}\vert 0\rangle_{\mbox{\rm\tiny conn.}}\
,
\end{eqnarray}
after integration over the solid angle in $q$-momentum space in the manner specified in
\cite{HPdR}. In Eq. (\ref{green}) $D_{\bar{s}q}=\overline{s}_Lq_R,\
G_{\bar{q}d}=\overline{q}_R d_L,\ L_{\bar{s}q}^{\mu}=\overline{s}_L\gamma^{\mu}q_L,\
R_{\bar{d}u}^{\alpha}=\overline{d}_R\gamma^{\alpha}u_R$, and similarly all the others.
One then recognizes the pair of fermion bilinears which make up the operators $Q_{6,4}$
in Eq. (\ref{penguins}) except that they are located at different space-time points. It
is the integral over $Q$ in Eq. (\ref{matching}) which brings them back to the same point
so that, in fact, the matching condition (\ref{matching}) is nothing but the statement
that in the four-quark Effective Lagrangian of Eq. (\ref{one}) one can only generate this
two-point correlator between $r_{\overline{d}u}^{\alpha}$ and $
r_{\overline{u}s}^{\beta}$ by inserting the combination $c_6
\mathcal{Q}_6+c_4\mathcal{Q}_4$. Since the same two-point correlator is proportional to
$g_{\underline{8}}$ in the language of the Chiral Effective Lagrangian, this explains the
origin of the condition (\ref{matching}).

It is clear that if one knew the functions $\mathcal{W}_{DG(LL)RR}$ one immediately would
be able to compute $g_{\underline{8}}$ through Eq. (\ref{matching}). The problem is of
course that these functions are not known. It is known, however, how they behave both at
low and at high values of $Q^2$ thanks to Chiral Perturbation Theory and the Operator
Product Expansion, respectively. So, if we can find a reliable way to interpolate between
the two, our job is done. At this point, large-$N_c$ comes to help. Since in the
large-$N_c$ limit both functions $\mathcal{W}_{DG(LL)RR}$ are made out of an infinity of
zero-width resonances, they are meromorphic functions. In other words, they only have
isolated poles, but no cut. Furthermore, they are order parameters of chiral symmetry and
would vanish were it not for its spontaneous breakdown. It follows, therefore, that
perturbation theory yields a vanishing result to all orders in $\alpha_s$ and, in
particular, there is no parton-model logarithm. With all these considerations, our choice
for the interpolator is then the most natural one, namely a meromorphic function with a
\emph{finite} number of poles. The position of these poles will be identified with the
resonance masses, and the unknown residues will be determined so that the interpolator
reproduces the low- and high-$Q^2$ expansions given by ChPT and the OPE. In mathematics
this is known as a rational approximant. Therefore, our interpolator constitutes an
approximation to the large-$N_c$ curve.

I would like to emphasize that this approximation \emph{is} systematic: the more terms in
the OPE and ChPT are known, the more resonances can be included in the
interpolator\footnote{The interpolator which matches just the first term in the OPE is
what we have sometimes called the ``Minimal Hadronic Approximation''\cite{Tempe} because
it is the simplest one which guarantees the correct short distance properties for the
matrix element.}. Although the solution to large-$N_c$ QCD is not known, it is plausible
that such an approximation may do a nice job. For one thing, we are interpolating a QCD
Green's function in the euclidean. Therefore one should not expect a lot of
``structure''; i.e. resonances do not show up as ``peaks'', unlike in the minkowski
region. For another thing, we interpolate in the gap between the regime governed by ChPT
and the OPE, so the gap does not seem very large! Of course, in the end, one will have to
judge by the results obtained.

What are the generic properties which one can expect for functions such as
$\mathcal{W}_{DG(LL)RR}(Q^2)$ ? Figure 1 is a schematic view of the expected typical
profile for one such generic QCD Green's function, here called $G(Q^2)$. As we can see,
at $Q^2=0$ the function reaches a value determined by a typical chiral parameter, such as
$F_0$ or $\langle\overline{\psi}\psi\rangle$, whereas at large $Q^2$ it falls off like an
inverse power of $Q^2$. The turning point is given by a typical resonance mass, of the
order of 1 GeV.

\vspace{-0.5cm}
\begin{figure}[htb]
\centering
\includegraphics[width=3in]{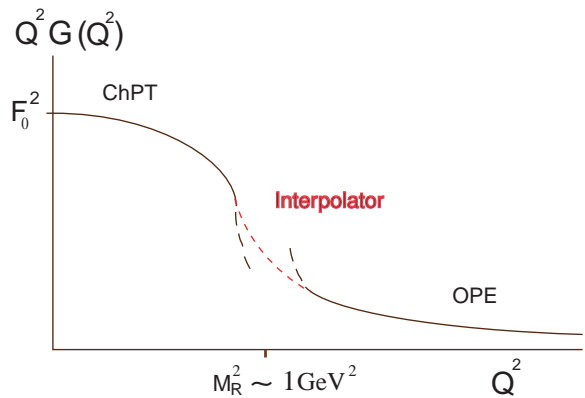}
 %\vspace{9pt} \framebox[55mm]{\rule[-21mm]{0mm}{43mm}}
 \vspace{-2cm}
 \caption{Schematic view of the QCD Green's function. The interpolation between ChPT and the OPE
 is represented by a (red) dashed curve .}
 \label{fig:1}
\end{figure}
\vspace{-0.5cm}

Let us come back to the matching condition (\ref{matching}). In trying to build some
intuition, it is important to keep track of all the different hadronic scales involved in
the problem since in QCD there is a very clear hierarchy, namely
\begin{equation}
    F_0\ll\langle\overline{\psi}\psi\rangle^{1/3}\ll M_R\sim 1\ \mathrm{GeV}\ ,
\end{equation}
and this determines the size of the final contribution. For instance, let us now estimate
the size of the unfactorized contributions relative to the factorized ones in the case of
$g_{\underline{8}}$. Equation (\ref{matching}) shows that the factorized contribution
from $Q_4$  is of order unity. In order to estimate its unfactorized contribution one can
use that the shape of $\mathcal{W}_{LLRR}(Q^2)$ is like that of $G(Q^2)$ in Fig. 1 and
therefore
\begin{equation}\label{estimate}
    \int dQ^2\ Q^2\ \mathcal{W}_{LLRR}(Q^2)\sim F_0^2 M_R^2\ ,
\end{equation}
up to logarithmic factors. Consequently the ratio of contributions
\begin{equation}\label{ratio}
    \frac{\mathrm{unfactorized}}{\mathrm{factorized}}\sim
    \frac{M_R^2}{16\pi^2F_0^2} \ ,
\end{equation}
which, although suppressed at large $N_c$ like $\mathcal{O}(1/N_c)$, is a number of order
unity. The important point is that this effect is very generic, and does not depend on
the particular form of the four-quark operator. For instance, despite appearances, the
contribution from $Q_6$ also obeys (\ref{ratio}): the factorized contribution is
proportional to $L_5 \langle\overline{\psi}\psi\rangle^2/F_0^6$, but the function
$\mathcal{W}_{DGRR}(Q^2)$ has the shape of $G(Q^2)$ in Fig. 1 multiplying its value at
the origin by $(L_5-5 L_3/2)\langle\overline{\psi}\psi\rangle^2/F_0^4$\cite{HPdR}. This
is a consequence of $Q_6$ being made up of scalar and pseudoscalar densities rather than
left and right currents.

One simple reason behind  the estimate given in (\ref{ratio})  is that unfactorized
contributions know about the scale $M_R\sim 1$ GeV whereas factorization is dominated by
chiral parameters, like $F_0$ or $\langle\overline{\psi}\psi\rangle$, whose typical
scales are smaller. Factorized contributions do not have ``access'' to the scale $M_R$.
In this sense it is something similar to the opening-up of a ``new channel'' in a
scattering process. Therefore, there is no reason to expect the next-to-next-to-leading
contribution to be even bigger, with the consequent breakdown of the large-$N_c$
expansion.  Since the large contribution is due to the coming into play of a new scale,
once all the scales have already appear in the problem there should be no more surprises.

There is another (independent) reason why, in the particular case of $Q_6$, one could
have expected large unfactorized corrections: the factorized contribution depends on
$L_5(\mu)$. Although the running of $L_5$ with $\mu$ is a $1/N_c$ effect, in fact, a
small change in scale such as going from the rho mass to 1 GeV  already changes the
factorized contribution by roughly a factor of 2 ! If one takes this as a naive estimate
for the $1/N_c$ corrections, one also concludes that it is not unnatural for these to be
large. Indeed, it was found that the unfactorized contributions were actually larger than
the factorized ones\cite{HPdR}. Within the context of a model, this was also found in
Refs. \cite{Hans}.

Using that the Wilson coefficients can be decomposed as\cite{anomdim}
\begin{equation}\label{wilson}
    C_i(\mu)\!=z_i(\mu)+\tau y_i(\mu),\
    \mathrm{with}\ \tau \!=-\frac{V^*_{ts}V_{td}}{V^*_{us}V_{ud}}\ ,
\end{equation}
the coupling constant $g_{\underline{8}}$ in Eq. (\ref{matching}) picks up an imaginary
part due to $\tau$. We obtain\cite{HPdR} $\left(\mathrm{Im}\tau= -6.0(5) \cdot
10^{-4}\cite{BurasII}\right)$
\begin{equation}\label{img8}
    \mathrm{Im}g_{\underline{8}}\simeq (3\pm 1)\times \mathrm{Im}\tau
     \ ,
\end{equation}
where the error has been estimated by varying the quark condensate (which is the source
of the biggest uncertainty), $\langle\overline{\psi}\psi\rangle^{1/3}(2\
\mathrm{GeV})=(0.240-0.260)\ GeV$. In Ref. \cite{KPdR} it was also obtained
that\footnote{See Ref. \cite{ewothers} for other analyses.}
\begin{equation}\label{gew}
    \mathrm{Im}\left(e^2g_{ew}\right)=
    (1.6 \pm 0.4)\cdot 10^{-6}\ \mathrm{GeV}^{6} \times \mathrm{Im}\tau\ .
\end{equation}
With (\ref{gew}) and (\ref{img8}) one can get to an estimate for
$\varepsilon'/\varepsilon$\cite{Buras}:
\begin{equation}\label{eps}
\frac{\varepsilon'}{\varepsilon}= \frac{\mathrm{Im}\left( V_{ts}^{*}V_{td}
\right)\mathrm{G_F}\omega}{2\vert\varepsilon\vert \big\vert\mathrm{Re} A_{0}\big\vert}\
\left[ P^{(0)}(1-\Omega_{\mbox{\rm\footnotesize IB}})-\frac{1}{\omega}P^{(2)} \right]
\end{equation}
where
\begin{equation}
P^{(0,2)}=\sum_{i=4,6,8}y_{i}(\mu)\langle (\pi\pi)_{0,2}\vert Q_{i}(\mu)\vert
K^{0}\rangle\ ,
\end{equation}
and $\Omega_{\mbox{\rm\footnotesize IB}}=0.16\pm 0.03$ \cite{oldPich} is a term induced
by isospin breaking. Using the above value for the condensate, $1/\omega=22.2$ and the
physical value for $\mathrm{Re}A_0$, we obtained in Ref. \cite{HPdR}
\begin{equation}\label{eps1}
    \frac{\varepsilon'}{\varepsilon}\simeq (2\pm 2) \times 10^{-3}\ .
\end{equation}
However, using the recent determination $\Omega_{\mbox{\rm\footnotesize IB}}=0.06\pm
0.08$ \cite{newPich} and including final state interactions as in Ref. \cite{SPP} one
obtains instead,
\begin{equation}\label{eps2}
 \frac{\varepsilon'}{\varepsilon}\simeq (5\pm 3) \times 10^{-3}\ .
\end{equation}
This is to be compared to the experimental number, i.e.
$\varepsilon'/\varepsilon=(1.66\pm 0.16)\times 10^{-3}$ \cite{exp}. The lesson is that,
due to the difference between the two terms in (\ref{eps}), small corrections to either
term get amplified making this observable extremely difficult to control. Matching the
level of precision of the experimental result is going to be a very arduous task.

 At present, the lattice determination of $\varepsilon'/\varepsilon$ \cite{lattice}
 is smaller than the experimental result by roughly a factor of three and differs in sign.
 Our previous considerations could be of help in
 understanding this situation. First of all, on the lattice one does
 not really have QCD, but a quenched
 version of it. This quenched version has some ghost quarks which,
 although spin one-half particles, are bosons. They are responsible
 for canceling the quark contribution in any internal quark loop. The presence of these
 ghosts dramatically
 changes the flavor symmetry of the
 theory  from the usual $SU(3)_L\times SU(3)_R$ to a graded $SU(3|3)_L\times
 SU(3|3)_R$\cite{BG}. As a consequence, an operator like $\mathcal{Q}_6$ in Eq. (\ref{penguins}) has no
 longer well-defined transformation properties under the flavor group but it can be
 decomposed as\cite{GP1}
\bqry\label{qchq6}
\mathcal{Q}_6&=&\frac{1}{2}\mathcal{Q}_6^{QS}+\mathcal{Q}_6^{QNS}\,,\\
\frac{1}{2}\mathcal{Q}_6^{QS}&=& 2(\sbar_L^\alpha\gamma_\mu d_L^\beta)
(\qbar_R^\beta\gamma^\mu q_R^\alpha+\tqbar_R^\beta\gamma^\mu \tq_R^\alpha)\,,
\nonumber\\
\mathcal{Q}_6^{QNS}&=& 2(\sbar_L^\alpha\gamma_\mu d_L^\beta) (\qbar_R^\beta\gamma^\mu
q_R^\alpha-\tqbar_R^\beta\gamma^\mu \tq_R^\alpha)\,, \nonumber \eqry where $q$ is summed
over $u,d,s$, and $\tq=\tu,\td,\ts$ are the bosonic (ghost) quarks. Whereas
$\mathcal{Q}_6^{QS}$ is a singlet under the quenched group $SU(3|3)_R$,
$\mathcal{Q}_6^{QNS}$ is not. This fact changes their corresponding chiral properties
completely. To leading order in ChPT, these operators may be represented by \cite{GP1}
\bqry \label{bosonization}\mathcal{Q}_6^{QS}&\to& -\alpha^{(8,1)}_{q1}\str(\Lambda
\mathcal{L}_\mu \mathcal{L}^\mu)\nonumber \\&& +\alpha^{(8,1)}_{q2}\str(2B_0\Lambda(U
M+MU^\dagger))\,,
\nonumber\\
\mathcal{Q}_6^{QNS}&\to& f^2\alpha^{NS}_q\str(\Lambda U\Nhat U^\dagger)\,,
\nonumber\\
\Nhat&=&\frac{1}{2}\,{\rm diag}(1,1,1,-1,-1,-1)\,, \eqry with $\qbar\Lambda q=\sbar d$,
$M$ the quark-mass matrix, and $\str$ is the so-called supertrace. $\Nhat$ represents the
non-singlet structure of $\mathcal{Q}_6^{QNS}$. Comparing Eqs. (\ref{ewcl}) and
(\ref{bosonization}), one sees that the couplings $\alpha^{(8,1)}_{q1,2}$ have a
corresponding counterpart in the unquenched theory (the weak mass term of
$\alpha^{(8,1)}_{q2}$ is not written in Eq. (\ref{ewcl})), but $\alpha^{NS}_q$ is a total
quenching artifact. Presently, $\alpha^{NS}_q$ is neglected in lattice simulations.

Notice that $\mathcal{Q}_6^{QNS}$ is very similar to the electroweak penguin
$\mathcal{Q}_8$ if one makes the replacement $Q\rightarrow \Nhat$. It follows that
$\alpha^{NS}_q$ must obey the same matching condition $g_{ew}$ does, and this leads
to\cite{KPdR}\footnote{Ref. \cite{MGSP} presents another derivation of this result.}
\bqry \alpha^{NS}_q&=&- F_0^2B_0^2 \left(1+O\left(\frac{1}{N_c^2}\right)\right)\,. \eqry
where $B_0=-\langle\overline{\psi}\psi\rangle/F_0^2$.

As to the operator $\mathcal{Q}_6^{QS}$, its contribution to $\alpha^{(8,1)}_{q1}$ is
also given by Eq. (\ref{matching}) in the leading-log approximation. One can neglect the
contribution from $\mathcal{Q}_4$ in this expression because it is numerically small and,
furthermore, $\mathcal{Q}_6^{QS}$ and $\mathcal{Q}_4$ no longer mix in the quenched
theory at leading log. One then concludes that the large unfactorized contributions which
were dominant in the unquenched theory are now vanishing due to the fact that $n_F$ is
effectively zero, as a consequence of the quark-ghost cancelation !\cite{MGSP}. One
predicts, therefore, that the result is given by the factorized contribution, which is a
\emph{much smaller} value for $\alpha^{(8,1)}_{q1}$ than in the unquenched theory : \bqry
\alpha^{(8,1)}_{q1}&=&-16L_5 F_0^2 B_0^2 \left(1+O\left(\frac{1}{N_c^2}\right)\right)\,.
\eqry Moreover, if one compares $\alpha^{NS}_q$ with $\alpha^{(8,1)}_{q1}$ one finds
\bqry \frac{\alpha^{NS}_q}{\alpha^{(8,1)}_{q1}} &=&\frac{1}{16L_5}\sim 60\,, \eqry where
we have used that $L_5\sim 10^{-3}$. This is not a small number and obviously questions
the validity of neglecting $\alpha^{NS}_q$ in lattice simulations of
$\varepsilon'/\varepsilon$ as is currently done.

I would like to conclude by coming back to unquenched QCD and turn to the $\Delta I=1/2$
rule. This rule can be expressed by the fact that $\mathrm{Re}\
g^{exp}_{\underline{8}}\simeq 3.3$ is much larger than $g^{exp}_{\underline{27}}\simeq
0.23$, after subtraction of the chiral corrections\cite{Ximo}. Any systematic framework
must also face the calculation of these values.

Although --due to mixing-- below the charm mass all operators contribute to $\mathrm{Re}\
g_{\underline{8}}$,  at $\mu=m_c$ only $\mathcal{Q}_{2,1}$ given by
\begin{eqnarray}\label{q12}
\mathcal{Q}_2 &=&4(\bar{s}_{L}\gamma^{\mu}u_{L}) (\bar{u}_{L}\gamma_{\mu}d_{L})\nonumber \\
 \mathcal{Q}_1 &=&4(\bar{s}_{L}\gamma^{\mu}d_{L})(\bar{u}_{L}\gamma_{\mu}u_{L})\
\end{eqnarray}
have non-vanishing contributions.  In fact, these contributions from $\mathcal{Q}_{1,2 }$
can be related (neglecting penguins\cite{PichdR}) to the coupling constant $g_{S=2}$
governing the local $K^0\leftrightarrow \overline{K}^0$ transition\cite{BK}. Now, our
calculations allow us to add to this the contributions coming from penguin
configurations. The point is that the penguin-like contribution from $\mathcal{Q}_2$
which is obtained by contracting the two $u$ quarks is the same as the unfactorized
contribution from $\mathcal{Q}_4$ in Eq. (\ref{matching}), via the replacement
$n_F\rightarrow 1$. One then obtains
\begin{equation}\label{Reg8}
    \mathrm{Re}g_{\underline{8}}=\!\!\!\!\underbrace{1.33\pm 0.40}_{(\mathcal{Q}_{2,1}
    \ \mathrm{non-penguin})} \!\!\!\!+
\underbrace{0.8\pm 0.4}_{(\mathcal{Q}_{2}\ \mathrm{penguin})}=2.1\pm 0.8\,.
\end{equation}
In spite of the large errors involved I find this result quite encouraging.

Furthermore, $g_{\underline{27}}$ can also be calculated. This is due to the celebrated
relation to $g_{S=2}$ and $B_K$\cite{Donoghue}. Using the results of Ref. \cite{BK}, one
finds
\begin{equation}\label{g27}
    g_{\underline{27}}=0.29\pm 0.12\  \Leftrightarrow \ \widehat{B}_K=0.36\pm 0.15\ ,
\end{equation}
in the chiral limit, with a nice agreement with $g^{exp}_{\underline{27}}$.

Other interesting applications of our approximation to large-$N_c$ QCD include the decays
$\pi^0\rightarrow e^+e^-$ and $\eta\rightarrow \mu^+\mu^-$\cite{mu}, hadronic vacuum
polarization\cite{Martina Franca} and the calculation of the hadronic light-by-light
contribution to $g_{\mu}-2$\cite{KN}.

\section*{Acknowledgements}

I am very thankful to M. Golterman, T. Hambye and E. de Rafael for a most enjoyable
collaboration which led to the completion of the work presented here. I am also thankful
to M. Knecht for lots of discussions. This talk was delivered at the  ``X Int. Conference
on Quantum Chromodynamics'', July 2003, Montpellier, France. I also wish to thank S.
Narison for inviting me to this pleasant conference. This work was supported by
CICYT-FEDER-FPA2002-00748, 2001 SGR00188 and by TMR EC-Contracts HPRN-CT-2002-00311
(EURIDICE).


\begin{thebibliography}{9}

\bibitem{Buras}G.~Buchalla et al.,
%``Weak Decays Beyond Leading Logarithms,''
Rev.\ Mod.\ Phys.\  {\bf 68} (1996) 1125.
%%CITATION = HEP-PH 9512380;%%


\bibitem{anomdim} A.~J.~Buras et al.,
%``Two loop anomalous dimension matrix for Delta S = 1 weak nonleptonic decays. 1. O(alpha-s**2),''
Nucl.\ Phys.\ B {\bf 400} (1993) 37;
%%CITATION = HEP-PH 9211304;%%
M.~Ciuchini et al.,
%``The Delta S = 1 effective Hamiltonian including next-to-leading order QCD and QED corrections,''
Nucl.\ Phys.\ B {\bf 415} (1994) 403.
%%CITATION = HEP-PH 9304257;%%
\bibitem{Tempe}S.~Peris,
%``Electroweak matrix elements at large N(c): Matching quarks to mesons,''
hep-ph/0204181;
%%CITATION = HEP-PH 0204181;%%
E.~de Rafael,
%``Analytic approaches to kaon physics,''
hep-ph/0210317.
%%CITATION = HEP-PH 0210317;%%




\bibitem{HPdR}
T.~Hambye et al.,
%``Delta(I) = 1/2 and epsilon'/epsilon in large-N(c) QCD,''
JHEP {\bf 0305}, 027 (2003).
%%CITATION = HEP-PH 0305104;%%

\bibitem{BurasII}A.~J.~Buras,
%``CP violation in B and K decays: 2003,''
hep-ph/0307203.
%%CITATION = HEP-PH 0307203;%%

\bibitem{Hans}T.~Hambye et al.,
%``Analysis of epsilon'/epsilon in the 1/N(c) expansion,''
Nucl.\ Phys.\ B {\bf 564} (2000) 391;
%%CITATION = HEP-PH 9906434;%%
J.~Bijnens at al., hep-ph/0309216
%%CITATION = HEP-PH 0309216;%%



\bibitem{KPdR} M.~Knecht et al.,
%``A critical reassessment of Q(7) and Q(8) matrix elements,''
Phys.\ Lett.\ B {\bf 508} (2001) 117;
%%CITATION = HEP-PH 0102017;%%
ibid.
%``Matrix elements of electroweak penguin operators in the 1/N(c)  expansion,''
Phys.\ Lett.\ B {\bf 457} (1999) 227.
%%CITATION = HEP-PH 9812471;%%


\bibitem{ewothers}See, e.g., V. Cirigliano, contribution to EPS-HEP2003, July 2003,
Aachen, Germany.





\bibitem{oldPich}G.~Ecker et al.,
%``pi0 eta mixing and CP violation,''
Phys.\ Lett.\ B {\bf 477} (2000) 88.
%%CITATION = HEP-PH 9912264;%%


\bibitem{newPich}V.~Cirigliano et al.,
%``Isospin violation in epsilon',''
hep-ph/0307030.
%%CITATION = HEP-PH 0307030;%%


\bibitem{SPP}I.~Scimemi et al.,
%``epsilon'/epsilon in the standard model,''
hep-ph/0111262.
%%CITATION = HEP-PH 0111262;%%


\bibitem{exp}
See, e.g., A.~Alavi-Harati et al.,
%``Measurements of direct CP violation, CPT symmetry, and other parameters in the neutral kaon system,''
Phys.\ Rev.\ D {\bf 67} (2003) 012005.
%%CITATION = HEP-EX 0208007;%%

\bibitem{lattice}T.~Blum et al.,
%``Kaon matrix elements and CP-violation from quenched lattice QCD. I: The  3-flavor case,''
hep-lat/0110075;
%%CITATION = HEP-LAT 0110075;%%
J.~I.~Noaki  et al.,
%``Calculation of non-leptonic kaon decay amplitudes from K $\to$ pi matrix  elements in quenched domain-wall QCD,''
Phys.\ Rev.\ D {\bf 68} (2003) 014501.
%%CITATION = HEP-LAT 0108013;%%

\bibitem{BG}C.~W.~Bernard at al.,
%``Chiral perturbation theory for the quenched approximation of QCD,''
Phys.\ Rev.\ D {\bf 46} (1992) 853.
%%CITATION = HEP-LAT 9204007;%%



\bibitem{GP1} M.~Golterman et al.,
%``Effects of quenching and partial quenching on penguin matrix elements,''
JHEP {\bf 0110}, 037 (2001).
%%CITATION = HEP-LAT 0108010;%%


\bibitem{MGSP}
M.~Golterman et al.,
%``Analytic estimates for penguin operators in quenched QCD,''
hep-lat/0306028.
%%CITATION = HEP-LAT 0306028;%%

\bibitem{Ximo} J.~Bijnens et al.,
%``K $\to$ 3pi decays in chiral perturbation theory,''
Nucl.\ Phys.\ B {\bf 648} (2003) 317;
%%CITATION = HEP-PH 0205341;%%
J.~Bijnens, et al.,
%``Obtaining K $\to$ pi pi from off-shell K $\to$ pi amplitudes,''
Nucl.\ Phys.\ B {\bf 521} (1998) 305, and refs. therein.
%%CITATION = HEP-PH 9801326;%%



\bibitem{PichdR}A.~Pich et al.,
%``Weak $K$--Amplitudes in the Chiral and $1/N_c$--Expansions,''
Phys.\ Lett.\ B {\bf 374} (1996) 186.
%%CITATION = HEP-PH 9511465;%%


\bibitem{BK} S.~Peris et al.,
%``K0 anti-K0 mixing in the 1/N(c) expansion,''
Phys.\ Lett.\ B {\bf 490} (2000) 213;
%%CITATION = HEP-PH 0006146;%%
O.~Cata et al.,
%``Long-distance dimension-eight operators in B(K),''
JHEP {\bf 0303} (2003) 060.
%%CITATION = HEP-PH 0303162;%%



\bibitem{Donoghue}J.~F.~Donoghue et al.,
%``The Delta S = 2 Matrix Element For K0 Anti-K0 Mixing,''
Phys.\ Lett.\ B {\bf 119} (1982) 412.
%%CITATION = PHLTA,B119,412;%%



\bibitem{mu}M.~Knecht et al.,
%``Decay of pseudoscalars into lepton pairs and large N(c) QCD,''
Phys.\ Rev.\ Lett.\  {\bf 83} (1999) 5230.
%%CITATION = HEP-PH 9908283;%%



\bibitem{Martina Franca}E.~de Rafael,
%``Large-N(c) QCD and low energy interactions,''
AIP Conf.\ Proc.\  {\bf 602} (2001) 14.
%%CITATION = HEP-PH 0110195;%%


\bibitem{KN}M.~Knecht et al.,
%``Hadronic light-by-light corrections to the muon g-2: The pion-pole  contribution,''
Phys.\ Rev.\ D {\bf 65} (2002) 073034.
%%CITATION = HEP-PH 0111058;%%




\end{thebibliography}
\end{document}